\documentclass[twocolumn]{aastex631}

\usepackage[english]{babel}
\usepackage{bbding}

\usepackage{amsmath}
\usepackage{amssymb}
\usepackage{graphicx}
\newcommand{\simlt}{\mathrel{\hbox{\rlap{\hbox{\lower4pt\hbox{$\sim$}}}\hbox{$<$}}}}

\newcommand{\mr}[1]{}
\newcommand{\hide}[1]{}

\shorttitle{Bursty Star Formation}
\shortauthors{Stiavelli \& Ricotti}

\graphicspath{{./}{Figs/}}

\begin{document}

% The minimal model for bursty star formation at $z>5$ that fits the data.
\title{How Bursty is Star Formation at $z>5?$}

\correspondingauthor{Massimo Stiavelli }
\email{mstiavel@stsci.edu}

\author[0000-0001-9935-6047]{Massimo Stiavelli}
\affiliation{Space Telescope Science Institute, 3700 San Martin Drive, Baltimore, MD 21218, USA}
\affiliation{Dept. of Physics \& Astronomy, Johns Hopkins University, Baltimore, MD 21218, USA}
%\affiliation{Dept. of Astronomy, University of Maryland, College Park, MD 20742, USA}

\author[0000-0003-4223-7324]{Massimo Ricotti}
\affiliation{Dept. of Astronomy, University of Maryland, College Park, MD 20742, USA}

%\maketitle

\begin{abstract}
Motivated by observational evidence from JWST and theoretical results from cosmological simulations, we use a simple parametric, phenomenological model to test to what extent bursty star formation with standard Initial Mass Function, no continuous star formation, no mergers, \mr{and no dust} can account for the observed properties in the $M_{UV}$ vs $M_*$ plane of galaxies at redshifts $z>5$. We find that the simplest model that fits the data has a quiescence period between bursts $\Delta t \sim 100$~Myrs and the stellar mass in each galaxy grows linearly as a function of time from $z=12$ to $z=5$ (i.e., repeated bursts in each galaxy produce approximately equal mass in stars). The distribution of burst masses across different galaxies follows a power-law $dN/dM_* \propto M_*^\alpha$ with slope $\alpha \sim -2$. 
%However, at higher redshifts we struggle to reproduce observations of stellar masses $>10^9$~M$_\odot$ in almost all models, suggesting that additional ingredients are necessary.
At $z>9-10$ the observed galaxy population typically had only one or two bursts of stars formation, hence the observed stellar masses at these redshifts (reaching $M_* \sim 10^{10}$~M$_\odot$), roughly represent the distribution of masses formed in one burst.  

%We show that our simple model naturally displays a double power-law luminosity function that reduces the number density of bright galaxies with $M_{UV}<-20$ at redshifts $z<9$. However, this reduction disappears at $z>9$, making bright galaxies at these redshifts more numerous, and hence more likely to be observed. This effect is due to the short cosmic time at $z>9$ when compared to the UV dimming timescale after a burst, that increase the galaxies duty cycle in the rest-frame UV.

\end{abstract}


\section{Introduction} \label{sec:intro}

%JWST intro. Star formation is bursty .... what are the implications?
With the advent of JWST evidence for a major role of bursty star formation in high redshift galaxies has gone up considerably, {\it e.g.} \citet{Adamo2024}. There is also evidence that the starburst and {\it normal} galaxy tracks merge together at high redshift and low galactic masses, {\it e.g.} \citet{Caputi2024}, \citet{Rinaldi2024}. Numerical simulations also provide support to the idea of bursty star formation as shown, {\it e.g.}, by \citet{sugimura2024,Pallottini2024,Garcia2025}. It should be noted that other groups have modelled stochastic or bursty star formation with different results. \mr{\citet{Gelli2024} adopted a mass-dependent stochasticity and found that they cannot fully reproduce observations \citep[see also][]{Sun2023, Ceverino2024, Ciesla2024}.}

On the basis of this evidence and these results, we have asked ourselves whether the entirety of star formation in low-mass galaxies could be happening in bursts without any significant continuous star formation, mergers, \mr{or dust attenuation}. If such a simple model can reproduce the observations, we would identify the model parameters that can be constrained robustly: for example, the quiescent time between bursts or the burst duration. Should our simple model fail to reproduce the observations we would learn, perhaps,  at what mass scales, it would break down or if bursty star formation alone is not enough.
To test this idea we have built a simple parametric phenomenological model of bursty star formation (see Sec.~\ref{sec:model}) and compared its predictions to observations in the $M_{UV}$ vs $M_*$ plane, also known as the star formation main sequence (SFMS), in galaxies at $z>5$ (Sec~\ref{sec:results}). We  will discuss the predicted luminosity function in Sec.~\ref{sec:LF} and draw our conclusions in Sec~\ref{sec:conclusions}.

\section{The Model}\label{sec:model}

In our toy model star formation happens in a series of instantaneous bursts labeled by the index {\rm i}. The mass contributed by the {\rm i}-th burst is a constant term plus a term proportional to the total stellar mass formed so far, namely:
\begin{equation}
    {\Delta} M_i = \xi + \eta M_{i-1}.
    \label{eq:one}
\end{equation}
The bursts are separated by a time $\Delta t$ which can be a constant or have a random dispersion. Unless otherwise stated we will use a constant $\Delta t$. 

The first burst takes place after redshift $z=30$ in a time interval chosen randomly between $0$ and $10 \Delta t$. 
The random start time for the sequence of bursts for each galaxy ensures that different galaxies are not {\it in phase} and a spread of UV luminosities at a given $M_*$ is observed. In practice, it also captures the diversity of halo masses at the initial redshift ($z=30$), simulating the fact that small-mass halos begin forming stars at later times than large-mass halos. 
%This ensures that the burst age will be random with respect to the sampling times and that some galaxies start forming stars later than others (presumably the ones hosted in smaller mass dark matter halos). 

For the first burst, {\rm i=1}, $M_0=0$ so $\Delta M_1 = \xi$. Therefore, the parameter $\xi$ can be interpreted as the distribution of masses formed in the first burst. In labeling the models, we measure $\xi$ in units of $10^6 M_\odot$.
%and $\Delta t$ in Gyrs.

Given that the elapsed time is proportional to the number of bursts (assuming constant $\Delta t$), Equation~(\ref{eq:one}) implies that the stellar mass in each galaxy increases linearly with time when $\eta = 0$ and exponentially with time when $\eta > 0$. Therefore, the value of $\eta$ describes a range of galaxies with different speeds of evolution for their stellar mass. Note that, as long as the interval between bursts is $\Delta t \leq$~100~Myrs, the galaxy rest-frame UV luminosity is dominated by the intensity and elapsed time from the last burst.

Since at a given redshift galaxies have a range of masses, we run models in which $\xi$ and $\eta$ are random values generated between predefined extremes, and we assume constant $\xi$ and $\eta$ only to explore the effects of changing these parameters.
For random values of $\xi$, we define a power-law slope $\alpha_\xi$ and generate $\xi$ uniformly for $\alpha_\xi=0$, uniformly in $log \xi$ for $\alpha_\xi=-1$ or uniformly in $\xi^{(1+\alpha_\xi)}$ for $\alpha_\xi \neq 0, -1$. We can generate $\eta$ in a similar way, keeping it constant or defining a coefficient $\alpha_\eta$ that characterizes the random distribution. We list in Table~\ref{tab:models} the basic properties of the subset of models that we discuss in the paper.

Galaxies evolve, burst after burst, until redshift 4 at which point the simulation ends. We define 21 sampling times uniformly spaced between redshift 12 and redshift 5 and "observe" each galaxy model at each of these times. For each "observation" we record the stellar mass and the 1500~\AA\ rest frame UV luminosity computed from the contribution of each past burst weighted through the burst age vs luminosity relation from the low-metallicity model ($Z=0.002$) with Salpeter IMF in STARBURST99 \citep{SB99}. 

The main comparison with JWST observations will be based on the $M_{UV}$ vs stellar mass distribution at $z>5$. We expected that at lower redshift other physical effects, such as continuous star formation, mergers, \mr{or dust attenuation may become important. The choice of not considering dust is justified by the widespread measurement of blue slopes in high-z galaxies \citep{Bouwens2023,Cullen2024,Dottorini2025,Franco2025,Donnan2025}. However, breaking down of this assumption would have a direct impact on our comparisons, {it e.g.} for a constant stellar mass dust attenuation in some objects would render them dimmer in the UV, thereby giving an effect that is degenerate with the interburst delay $\Delta t$. A mass-dependent dust attention as suggested by some observations \citep[e.g.][]{Dottorini2025,Franco2025} would further complicate the analysis and interpretation.}

In the figures and in our analysis, we consider redshift intervals containing the same number of "observations" and characterized by the same cosmic volume so that, after galaxies have had their first burst, the number of galaxies per cosmic volume remains constant. However, given that the first burst happens randomly in the time interval $0-10 \Delta t$ from $z=30$, the number of galaxies is not strictly constant because some galaxies, not having experienced their first burst yet, are dark.
Thus, aside from the presence of some dark galaxies (or galaxies below the detection limit), the evolution we see is effectively produced by the evolution of the luminosity function rather than the evolution of the galaxy number density.

 We have also explored the case where the first burst occurs randomly within the interval $0-\Delta t$ from $z=30$, so that effectively all galaxies form stars from the beginning of the simulation (at $z>12$), giving a pure luminosity evolution as the total number of luminous galaxies between z=5 and z=12 is constant ({\it e.g.} Model 1ND in Table~\ref{tab:models}).  \mr{This choice of initial burst timing gives us more galaxies at the highest redshift 
 %but consistently under-predicts faint (from $M_{UV} \sim -20$ to $-18$), low-mass ($M_* \sim 10^7-10^8$~M$_\odot$) galaxies in the redshift range $z\sim 5-6.3$. 
 but it seems to be able to reproduce properties similar to the other models when changing the parameters. Given that such a high level of synchronization of the beginning of star formation across many galaxies with a range of luminosities seems unrealistic, 
 %For this reason 
 we do not further consider models with no, or only small ($\leq \Delta t$), delay in star formation.} 

We can include an observability condition by requiring, {\it e.g.}, that only objects with apparent AB magnitude brighter than some value $m_{lim}$ are observable. In the following we adopt $m_{lim} = 30$  in a narrow observer frame band located at the rest frame 1500~\AA\ wavelength. Only objects brighter than this limit will be shown in the plots.

\section{Results} \label{sec:results}

%\mr{We should mention earlier that we focus on reproducing the $M_{UV}$ vs stellar mass relationship from JWST observations at $z>5$.}
In this paper, we are focusing our comparison on $M_{UV}$ vs stellar mass ($M_*$) rather than the Star Formation Rate (SFR) vs $M_*$, as the former quantity is in principle directly observable and easily determined from the model. Star Formation Rate is usually determined by photometric fits that include multiple bands and it requires us to adopt a smoothing model for what would otherwise be the sum of delta-functions. 

\subsection{Minimal Toy Models}\label{ssec:obs}

\begin{figure}[t]
%\centering
%\includegraphics[trim={0 6cm 0 6cm},clip,width=1.0\linewidth]{Figs/FigConstant.pdf}
%\includegraphics[width=1.0\linewidth]{Figs/FigConstant.png}
%\includegraphics[trim={0 6cm 0 6cm},clip,width=1.0\linewidth]{Figs/FigConstantBlock.pdf}
\includegraphics[width=1.0\linewidth]{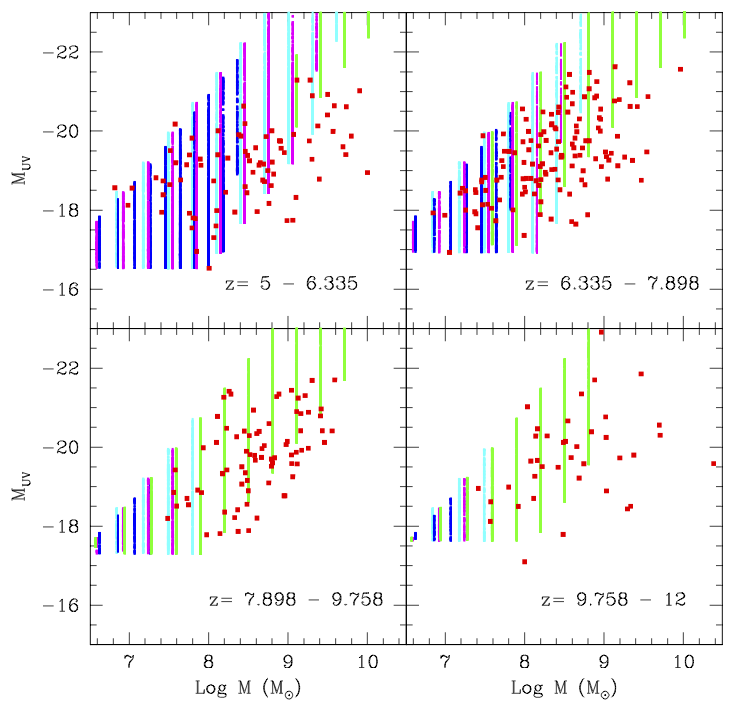}
\caption{\label{fig:constant} The cyan points show $M_{UV}$ vs $M_\star$ for a model with $\xi = 1$, $\eta = 1$, and $\Delta t = 100 Myrs$ (Model A in Table~\ref{tab:models}). Blue is for $\xi=1$, $\eta=0.5$ (Model B, displaced by -0.05 in log M for visibility purposes), and magenta for $\xi=0.5$ and $\eta=1$ (Model C, displaced by +0.05 in log M). Green is for with $\xi = 1$, $\eta = 1$, and $\Delta t = 50Myrs$ (Model D). We are only plotting models with apparent magnitude  $m_{UV} \leq 30$ to simulate an observational limit. Each vertical line corresponds to a given number of bursts and the number of bursts is related to redshift but is not a one-to-one function of redshift because of the random distribution of starting redshifts. The vertical extent of each line is related to the burst length $\Delta t$. The observational points in red are from \citet{Morishita2024}. 
%The vertical dashed line indicates the maximum mass achievable with $\eta =1$ and a number of bursts set by Hubble time divided by burst separation.
}
\end{figure}

First, we test minimal models in which $\xi$, $\eta$ and $\Delta t$ are constant. These models are not realistic but we show them to illustrate the effects of varying each of the 3 free parameters in the model.
For any choice of these parameters the models appear as a number of vertical lines in the $M_{UV}$ vs $M_\star$ plane. Fig.~\ref{fig:constant} shows the distribution in our model (vertical lines) compared to the \citet{Morishita2024} sample of galaxies at $z>5$. \mr{We have chosen this sample for comparison because it provides stellar masses and absolute UV luminosities, and we did not find better publicly available samples suited for our comparison. Unfortunately, the sample is obtained by combining different data sets with uneven selection criteria and a mix of spectroscopic and photometric redshifts. These limitations make it less suitable for a detailed comparison but we think it provides a good indication of what the envelope of galaxy properties is at these redshifts. The heterogeneity and limitations of the observational sample makes it suitable only for a qualitative comparison with the models. For this reason and for the sake of simplicity, we compare the observed UV magnitude to the value at 1500~\AA\ rather than calculating it for specific JWST filters.}

Each vertical line, at a given mass, corresponds to a given number of bursts. A summary of the effects of varying each parameter while keeping the others constant is given below. \\
i) {\bf Quiescence period $\Delta t$:} The vertical extent of the lines depends on $\Delta t$. Larger values of $\Delta t$ introduce a larger luminosity scatter by enabling the most recent burst to age as much as $\Delta t$. Smaller $\Delta t$ correspond to less aging and thus less dimming in the UV and, at the same time, galaxies with larger stellar mass as the number of bursts increases with decreasing $\Delta t$.  The implications of this result is that for each choice of parameters we only see a relatively narrow range of discrete masses at each redshift. This is the main rationale for considering random distributions of $\xi$ and/or $\eta$, in addition to also being a more physically motivated choice given the range of halo masses at any redshift.
%\cite{greenwade93}
%\begin{figure*}
%\centering
%\includegraphics[width=1.0\linewidth]{Figs/FigConstantOneRow.png}
%\caption{\label{fig:constant} The cyan points show $M_{UV}$ vs $M_\star$ for a model with $\xi = 1$, $\eta = 1$, and $\Delta t = 0.1$. Blue is for $\xi=1$, $\eta=0.5$ (displaced by -0.05 in log M for visibility purposes), and magenta for $\xi=0.5$ and $\eta=1$ (displaced by +0.05 in log M). Green is for with $\xi = 1$, $\eta = 1$, and $\Delta t = 0.05$. We are only plotting models with apparent magnitude  $m_{UV} \leq 30$ to simulate an observational limit. Each vertical line corresponds to a given number of bursts and the number of bursts is related to redshift but is not a one-to-one function of redshift because of the random distribution of starting redshifts. The vertical extent of each line is related to the burst length $\Delta t$. The observational points in red are from \citet{Morishita2024}. The vertical dashed line indicates the maximum mass achievable with $\eta =1$ and a number of bursts set by Hubble time divided by burst separation.}
%\end{figure*}
%\section{Comparison with Observations}\label{sec:comparison}

ii) {\bf Strength of first burst $\xi$:} Comparison of the model with the data in Fig.~\ref{fig:constant} shows that constant $\xi$, $\eta$ models do not seem to span the whole range of observed masses while also potentially over-predicting the observed luminosity. The latter effect is due to the fact that for $\eta=1$, models with mass larger than $10^8$ M$_\odot$ can have extremely bright bursts exceeding $10^8$ M$_\odot$ of new stars. 

iii) {\bf Growth rate of the burst $\eta$ :} Models with $\eta=0$ (linear mass growth rate) require values of $\xi$ generally larger than in models with $\eta>0$. However, both choices can be tuned to reproduce the range of $M_*$ observed in the redshift range $z=5-12$. 
%\mr{I am actually surprised that the liner model does not work better than the exponential given that the range of $M_*$ at $z>10$ is about the same as at $z\sim 5$. Where are all the galaxies that had $M=10^{10}$ at $z=10$ at $z=5$? If the growht is exponential they should be huge. A linear groth would only make them a few 100 times more massive.}

In all models the main challenge is to produce galaxies with the largest $M_*$ without making too many overly bright galaxies (with $M_{UV}<-22$).
We will see that we found two solutions to this problem: 

1) Increasing the duration of each burst. This allows each burst to produce a large mass in stars without making overly luminous galaxies in the rest-frame UV, as the formation of short-lived massive stars is spread over a time larger than their life on the main sequence, setting a maximum limit to their instantaneous number.

2) Assuming a distribution for $\xi$ that makes the number of galaxies producing large bursts in stars relatively rare, hence more unlikely to be observed at their peak luminosity. Nevertheless these galaxies exist, grow to have a large stellar mass, but are typically observed at fainter UV brightness than at their peak luminosity (i.e., after a major burst of star formation).

\subsection{More realistic models}

\begin{figure}
%\centering
\includegraphics[trim={0 6cm 0 6cm},clip,width=1.0\linewidth]{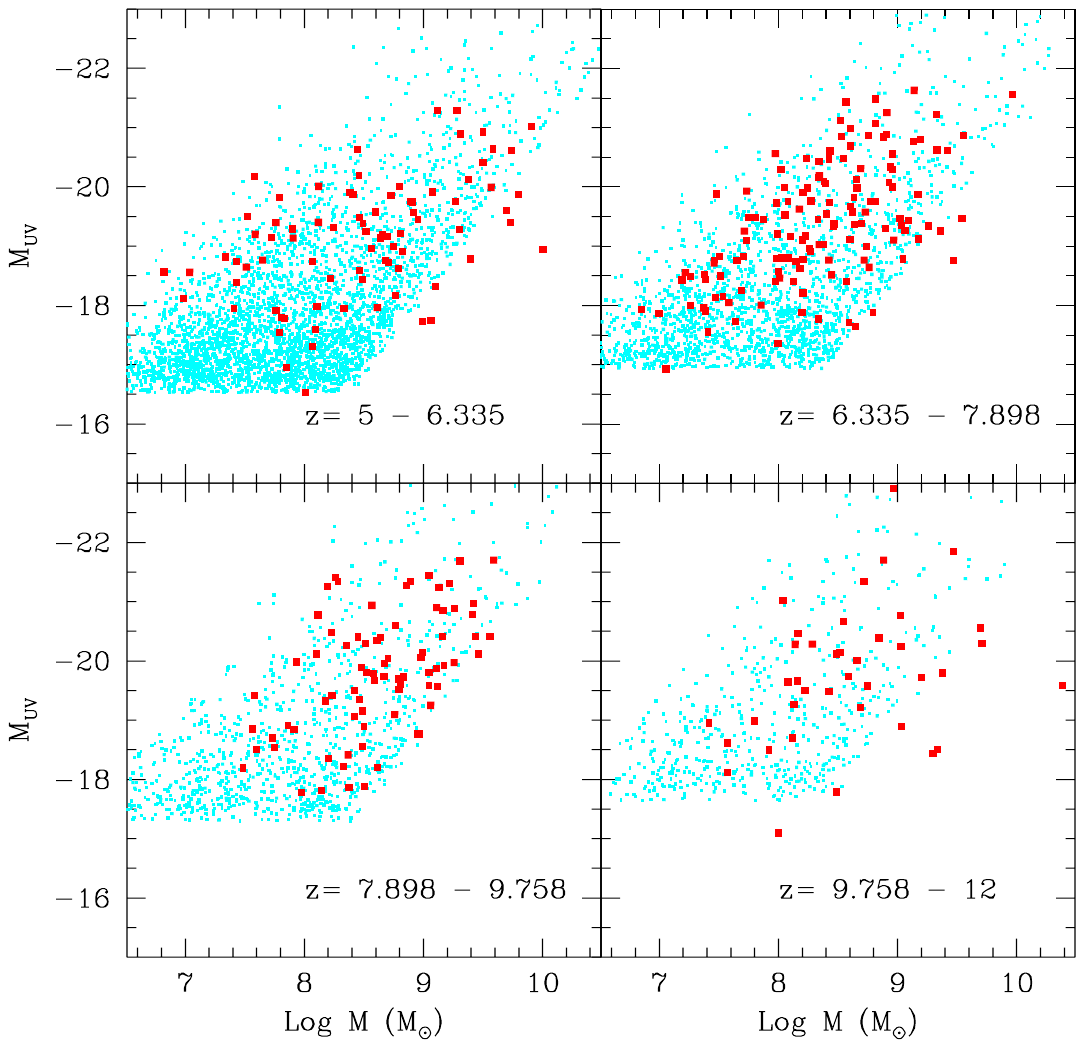}
\caption{\label{fig:MUV_M_nw1} $M_{UV}$ vs $M_\star$ for the galaxies in \citet{Morishita2024} compared to a model with a random distribution of $\xi$ and $\eta=0$ as described in the text (Model 20 in Table~\ref{tab:models}). The redshift bins in each panel have the same cosmic volume.}
\end{figure}

On the basis of the inadequacy of models with constant $\xi$ and $\eta$, let's consider now a model where $\eta=0$, $\Delta t = 100$ Myrs, and where $\xi$, i.e. the mass-independent strength of a burst, is randomly assigned to each galaxy with a uniform power-law distribution with slope $\alpha_\xi = -1.8$ and with $\xi_{min} = 1$ and $\xi_{max} = 5000$.  The resulting distribution is given in Fig.~\ref{fig:MUV_M_nw1} (see also Model 20 in Table~\ref{tab:models}). This particular model seems to reproduce the observed distribution at all redshift. Models with lower values of $\xi_{max}$, as {\it e.g.} Model 1 in  Table~\ref{tab:models}) are unable to produce the high masses seen at $z > 10$.

Assuming constant $\xi=1$ and a random $\eta$ uniformly extracted in, {\it e.g.}, the range 0.01-3 we would obtain very high luminosities due to very intense bursts (Model 9M in Table~\ref{tab:models}). To maintain the match with observations we need to effectively increase the duration of each burst. We do this, within our simple scheme, by introducing a maximum instantaneous burst mass. Essentially, we split the desired burst into multiple smaller bursts occurring one after the other with a 5 Myrs separation. This was not needed for the model above because the overall probability of intense bursts was lower.

We also consider models where both $\xi$ and $\eta$ are randomly distributed with $\alpha_\xi = -1.5$ and $\alpha_\eta=-1.5$ ({\it e.g.} Model 5 in Table~\ref{tab:models}). Even these models  can reproduce the range of masses and luminosities that are observed but tend to under-predict the most luminous objects at $z\geq8$ unless $\xi_{max} > 1000$. Thus, considering $\eta \neq 0$ does not allow us to avoid large values of $\xi_{max}$.

\begin{figure}
%\centering
\includegraphics[trim={0 6cm 0 6cm},clip,width=1.0\linewidth]{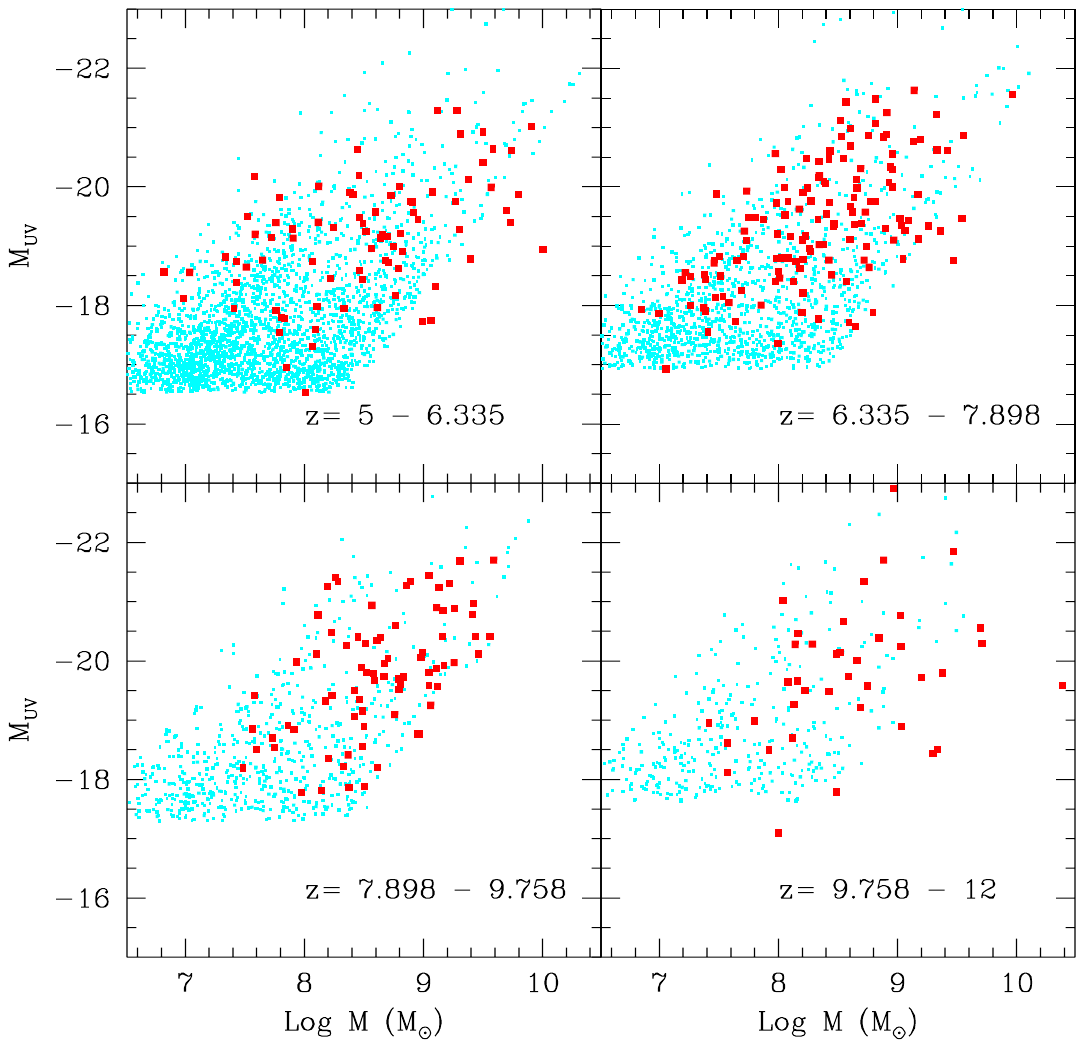}
\caption{\label{fig:best} $M_{UV}$ vs $M_\star$ for the galaxies in \citet{Morishita2024} compared to a model with $\eta = 0$ and  $\xi$ randomly distributed with $\alpha_\xi = -2$. Details are described in the text (Model 19 in Table~\ref{tab:models}). The slope of luminosity function for this model is compatible with the observed one.
}
\end{figure}
We have verified that the results discussed above would not change if the interburst spacing $\Delta t$ was given a random dispersion instead of being fixed to a given value.

\section{Luminosity Function slope}\label{sec:LF}

\begin{figure}
%\centering
\includegraphics[trim={0 6cm 0 6cm},clip,width=1\linewidth]{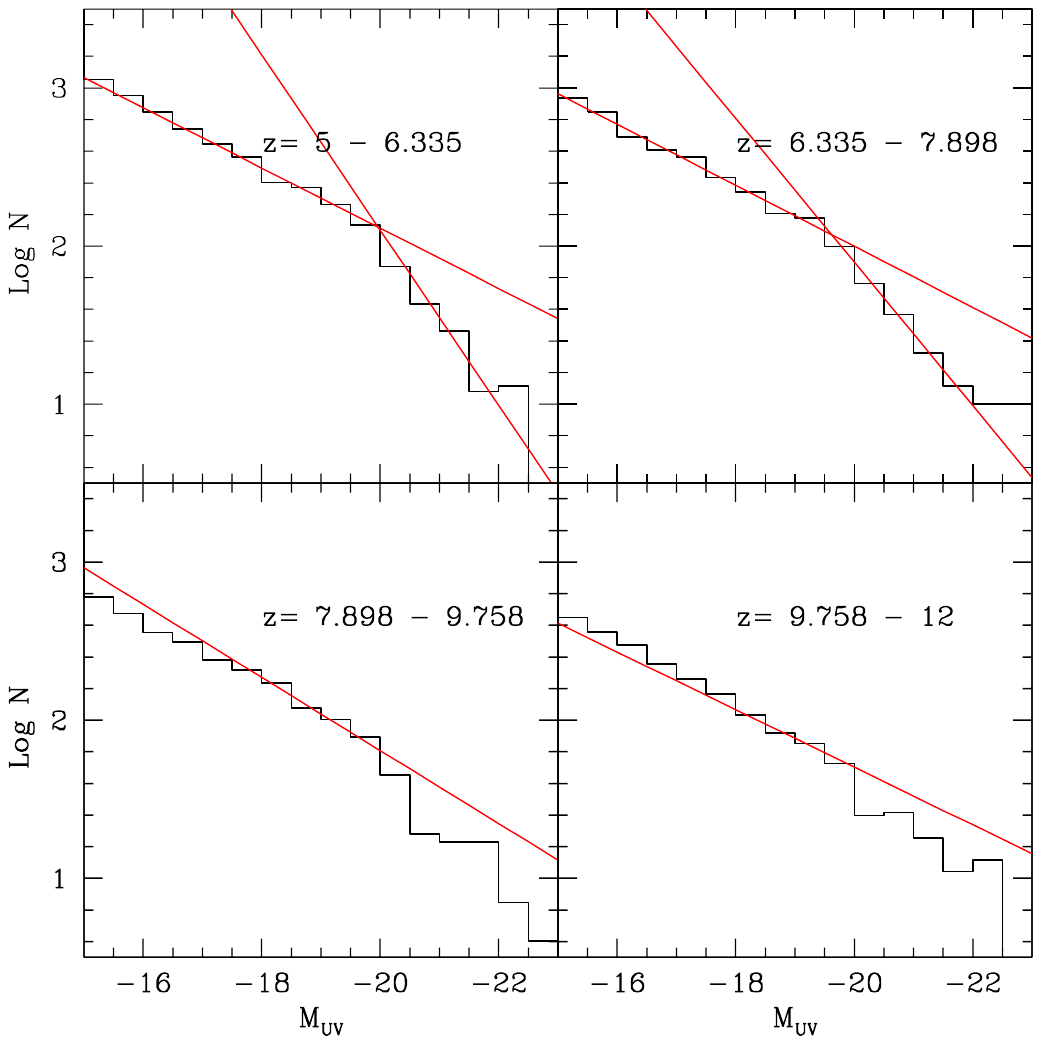}
\caption{\label{fig:LF_nw1} log counts vs $M_{UV}$  for the models with $\eta=0$ and a random power-law distribution of $\xi$ (Model 1). The slope break around $M_{UV}\simeq-21$ is due to the adopted $\Delta t = 0.1$ and the dimming of a burst over the time $\Delta t$. The red lines show a broken power low fit to the LF with $\alpha_{faint}=-0.2$ and $\alpha_{bright}=-0.5$ at $z=5-6$, $\alpha_{faint}=-0.19$ and $\alpha_{bright}=-0.27$ at $z=6-8$, and $\alpha_{faint}=-0.2$ at higher z.}
\end{figure}

From our models we can derive the luminosity function (LF) at different redshifts. For models where $\eta$ is a constant and $\xi$ is a random power-law distribution with slope $\alpha_\xi$, naively we expect a slope of the LF as a function of luminosity identifical to the slope in $\xi$: i.e., $\alpha_{LF} = \alpha_\xi$. Therefore the LF slope as a function of UV magnitude is $\alpha_{LF,mag} = 0.4 (\alpha_\xi + 1)$, obtained converting from luminosity to magnitude distributions. 
In Fig.~\ref{fig:LF_nw1} we show the LF for a model with $\eta=0$ and a random distribution of $\xi$ with slope $\alpha_\xi = -1.5$ (aka Model 1). No magnitude cutoff has been applied here to allow for a better fit. We find that the faint end slope $\alpha_{LF,mag}\approx -0.2$ of the LF agrees with the slope predicted above (dictated by the slope imposed for $\xi$, $\alpha_\xi=-1.5$). 

At redshifts $z<8$ the LF is well described by a broken power-law, even though our $\xi$ distribution is a single power-law with slope that matches the faint-end of the LF. The slope of the power-law at the bright end is steeper than at the faint end. We explain this effect as produced by the duty cycle of the bursts in the rest-frame UV: the number of galaxies that are observed at a given UV luminosity is proportional to the fraction of time they spend at that given luminosity (i.e., the duty cycle). Since, for an instantaneous burst, the UV luminosity drops very rapidly with elapsed time from the burst, the time spent at the peak luminosity is short, while the time spent at fainter luminosity is longer. Therefore, the number of bright galaxies is suppressed more strongly than the number of faint galaxies, producing a steeper drop at the bright end when compared to the input power-law of the bursts. 
In other words, the break is not caused by the assumed burst distribution but by the fact that a burst can be randomly observed at maximum brightness or after a time $\Delta t$ has elapsed ({\it i.e.} just before the next burst occurs). Reducing the value of $\Delta t$ moves the break to brighter magnitudes, and vice-versa. 

It is interesting that this broken power law behavior is unrelated to the exponential cut off of the dark matter halos mass function (which we do not consider here) but it's simply due to the imposed burst strength power-law and the dimming due to the inter-burst spacing. Note that in our model the break between the two power-laws is independent of redshift as long as the inter-burst spacing is also independent of redshift. 

In order to reproduce the observed luminosity function $\alpha_{LF} \simeq -2$ we need $\alpha_{\xi} \simeq -2$. For such slopes the break discussed above is not readily visible. \mr{However, our models with such slope do not show evidence for a steeper slope at brighter magnitudes in contrast to what seems suggested by observations \citep{Donnan2024}. In our bursty models to obtain a steeper slope at the bright end of the LF requires a slower dimming of the galaxy after the burst than obtained assuming a Salpeter IMF and instantaneous star formation. Hence we speculate that assuming a bottom heavy IMF or longer duration of the burst could improve the agreement with the observed LF at the bright end.}

\subsection{Minimal model that fits the data}

It turns out that the model that best fits the data is also the simplest. This is Model 19, which has linear growth $\eta=0$, burst spacing $\Delta t=100$~Myr, and a power-law distribution for $\xi$  in the range $1<\xi<10^4$ with slope -2 (see Fig.~\ref{fig:best}).

We note that the distribution of stellar masses at $z>10$ roughly represents the distribution of initial masses $\xi$ of the first burst. Hence, we chose a $\xi$ that fits the high-z galaxies. Since the \mr{upper envelope} of the  observed stellar masses do not increase substantially from redshift $z\sim 10$ to $z\sim 5$, a model with galaxies having exponential growth of their stellar mass does not match the data.

It is worth noting that cosmological zoom-in simulations of galaxies at $z>8$, in which the resolution is sufficient to resolve the formation of compact star cluster, show a bursty SF with $\Delta t \sim 100$~Myr and roughly equal strength of repeated starburst \citep{Garcia2025}, in good agreement with our minimal model.

\section{Discussion and conclusions}\label{sec:conclusions}

We find that a simple toy model where we assign to each galaxy two burst parameters and define a burst time separation for all objects can reproduce the distribution of galaxies in the Mass-UV luminosity plane at $z\geq5$. In addition the model the best reproduces the observation is even simpler: it has a linear growth of the stellar mass in each galaxy ($\eta=0$) and a constant time between bursts of 100~Myr. 

\mr{During the preparation of this paper we became aware of the results by \citet{Mitsu2026} that on the basis of analysis of UNCOVER data find that observations are compatible with bursty star formation independent of redshift. This is broadly compatible with our findings even though the details of the bursts are different.}

Below is a summary of our findings:
\begin{enumerate}
\item \mr{All our models assumes a standard Initial Mass Function and they don't include continuous star formation, merging, or dust attenuation}. Given these assumptions, the model that best reproduces the observations has a linear growth of the stellar mass in each galaxy ($\eta=0$) and a constant time between bursts of 100~Myr.
\item The distribution of initial masses of the galaxies (during their first burst) is a power law with slope $\alpha_\xi \sim -2$, within a minimum and maximum mass of (10$^6$ ,10$^{10}$)~M$_\odot$.  \mr{Perhaps this is a coincidence, but we note that this slope is compatible with what is seen locally for the LF of star clusters \citep{PZ2010,AdamoBastian2015}}.
This distribution also represents the distribution of stellar masses of observed galaxies at $z>10$. The largest stellar mass at these redshift is $M_* \sim 10^{10}$~M$_\odot$. This is a large value but not physically unrealistic for dark matter halos of mass $M\sim 10^{11}-10^{12}$~M$_\odot$, which are rare at $z>10$. Here we take a purely empirical approach, and we do not speculate on whether it is possible to form these many stars in a single burst of star formation. Several previous papers \citep{Dekel2023, Ferrara2024, Andalman2025} have proposed mechanisms that may be at play in the high redshift universe that may facilitate such large starbursts. 

\item  {The galaxy luminosity function at high redshift appears well fitted by a double power law \citep[e.g.][]{Donnan2024} while our models reproducing naturally the faint end slope do now show such feature. In contrast, our models with shallower fain-end slope display a steeper slope  at high luminosity - but not as steeps as observed - due to burst aging effects. The inability of our model to reproduce the steep bright end slopes could be due to the fact that we do not incorporate possible effects related to the exponential cut off in the number density of massive halos in the halo mass function, or other effects related to burst duration and aging, or changes in the IMF in brighter galaxies. Exploring this effect further goes beyond the purpose of this paper.}
\end{enumerate}
%The toy model assumes a standard Initial Mass Function and includes neither continuous star formation nor merging. Thus, it is not entirely surprising that additional physical processes may be necessary for galaxies of mass $10^9 M_\odot$ or higher. However, it is interesting that the observations of lower mass objects could be explained solely by simple bursty star formation.

%Most of our models struggle to predict the most massive galaxies observed at $z\geq8$. This cannot be entirely explained by the rarity of these objects. Indeed, if we increase the size of our Monte Carlo, a few more objects would populate the $z\geq8$ panels but the number of objects in the lower redshift panels would also continue to increase. 
 
%Thus, in agreement with common interpretations in the literature \citep{Adamo2024} reproducing observations would require either a significant quenching mechanism preventing further bursts at $z\leq 8$, contribution by dust dimming galaxies at lower redshift compared to those at the highest redshift, or a redshift dependent stellar Initial Mass Function (more top heavy at high redshift).

\begin{deluxetable*}{cccccc|ccc|l}
\tablecaption{
Summary of model properties
}
\tablehead{
\colhead{ID} & \colhead{$\xi$} & \colhead{$\eta$} & \colhead{$\Delta t$} & 
\colhead{1st burst} & \colhead{Burst} &
\multicolumn{3}{|c}{Model vs Data} & \multicolumn{1}{|c}{$Notes$}\\
\colhead{} & \colhead{} & \colhead{} & \colhead{(Myrs)} & 
\colhead{delay (in $\Delta t$)} & \colhead{splitting} &
\multicolumn{1}{|c}{[$M_*$ @ $z>9.76$]} & \colhead{[$M_{UV}<-22$]} & \colhead{Verdict} & \multicolumn{1}{|c}{}
}
\startdata
A & 1   & 1   & 100 & 10 & N & 87\% & 45\% &\XSolidBrush/\XSolidBrush & exp. SF\\
B & 1   & 0.5 & 100 & 10 & N & 95\%&  6\% & \XSolidBrush/\Checkmark & \\
C & 0.5 & 1   & 100 & 10 & N & 94\%& 39\% & \XSolidBrush/\XSolidBrush& dep. on $\xi$\\
D &  1  & 1   & 50  & 10 & N & 57\%& 48\% & \XSolidBrush/\XSolidBrush& dep. on $\Delta t$\\
E &  1  & 0   & 100 & 10 & N & 100\% & 0\% & \XSolidBrush/\Checkmark & uniform SF\\
1 & (0.17,480,-1.5) & 0 & 100 & 10 & N & 32\% & 18\% & \XSolidBrush/\Checkmark & \\
1ND & (0.17,480,-1.5) & 0 & 100 & 1 & N & 24\% & 12\% & \Checkmark/\Checkmark & LF too flat \\
5 & (0.17,480,-1.5)   & (0,3,-1.5) & 100 & 10 & N & 28\% & 47\% & \Checkmark/\XSolidBrush& \\
8 & (0.17, 480, -1.5) & 1 & 100  & 10 & N & 24\%& 52\% & \Checkmark/\XSolidBrush & \\
8M & (0.17, 480, -1.5) & 1 & 100  & 10 & Y & 46\%& 7\% & \XSolidBrush/\Checkmark & \\
9M & 1      & (0,3,0) & 100  & 10 & Y & 73\%& 6\% & \XSolidBrush/\Checkmark & \\
10 & (0.17,480,-1.5) & (0, 3, 0) & 100 & 10 & N & 21\% & 55\% &  \Checkmark/\XSolidBrush & \\
11 & (0.17,2000,-1.5) & 0 & 100 & 10 & N & 20\% & 29\% &  \Checkmark/\Checkmark & LF too flat\\
12 & (0.17,2000,-1.8) & 0 & 100 & 10 & N & 40\% & 25\% &  \XSolidBrush/\Checkmark &\\
13 & (0.17,5000,-1.8) & 0 & 100 & 10 & N & 37\% & 30\% &  \XSolidBrush/\XSolidBrush & \\
18 & (0.17,10000,-2) & 0 & 100 & 10 & N & 62\% & 9\% &  \XSolidBrush/\Checkmark &\\
19 & (1,10000,-2) & 0 & 100 & 10 & N & 28\% & 21\% &  \Checkmark/\Checkmark & best model\\
20 & (1,5000,-1.8) & 0 & 100 & 10 & N & 21\% & 28\% & \Checkmark/\Checkmark & good \\
\enddata
\tablecomments{
A more detailed description of the parameters is in the text. In the $\xi$ and $\eta$ columns we either give one value or three. When one value is given, the parameter is kept constant at that value. When three are given, the parameter is varied between the first two values using a power-law slope given by the third value. When burst splitting is enabled we split any burst creating more than 60M M$_\odot$ into multiple bursts separated by 5 Myrs.
}\label{tab:models}
\end{deluxetable*}

\begin{acknowledgments}

We thank the anonymous referee for comments that helped improved the paper. MS acknowledges the hospitality of the University of Maryland - College Park during the preparation of this paper as well as support for this work under NASA grant 80NSSC22K1294.

\end{acknowledgments}

%% To help institutions obtain information on the effectiveness of their 
%% telescopes the AAS Journals has created a group of keywords for telescope 
%% facilities.
%
%% Following the acknowledgments section, use the following syntax and the
%% \facility{} or \facilities{} macros to list the keywords of facilities used 
%% in the research for the paper.  Each keyword is check against the master 
%% list during copy editing.  Individual instruments can be provided in 
%% parentheses, after the keyword, but they are not verified.

\vspace{5mm}
%\facilities{JWST(NIRSpec)}

%% Similar to \facility{}, there is the optional \software command to allow 
%% authors a place to specify which programs were used during the creation of 
%% the manuscript. Authors should list each code and include either a
%% citation or url to the code inside ()s when available.

%\software{
%Astropy \citep{astropy13,astropy18,astropy22}, bbpn \citep{bbpn}, EAzY \citep{brammer08}, EMCEE \citep{foreman13}, gsf \citep{morishita19}, numpy \citep{numpy}, python-fsps \citep{foreman14}, JWST pipeline \citep{jwst},
%msaexp \citep{brammer23}.
        %astropy \citep{2013A&A...558A..33A,2018AJ....156..123A},  
        %  Cloudy \citep{2013RMxAA..49..137F}, 
        %  Source Extractor \citep{1996A&AS..117..393B}
%          }

\bibliography{sample}{}
\bibliographystyle{aasjournal}

\end{document}